\documentclass[%
reprint,
superscriptaddress,
amsmath,amssymb,
aps,
]{revtex4-2}

\usepackage{graphicx}
\usepackage{dcolumn}
\usepackage{bm}
\usepackage{verbatim}   
\usepackage{physics}
\usepackage{amsmath}
\usepackage{xcolor}
\usepackage{float}
\usepackage{soul}   
\usepackage{hyperref}
\usepackage[normalem]{ulem}

\begin{document}

\preprint{APS}

\title{Matter-wave interferometers with trapped strongly interacting Feshbach molecules}

\author{Chen Li}
\email{chen.li@tuwien.ac.at}
\affiliation{Vienna Center for Quantum Science and Technology, Atominstitut, TU Wien, Stadionallee 2, 1020 Vienna, Austria}

\author{Qi Liang}
\affiliation{Vienna Center for Quantum Science and Technology, Atominstitut, TU Wien, Stadionallee 2, 1020 Vienna, Austria}

\author{Pradyumna Paranjape}
\affiliation{Vienna Center for Quantum Science and Technology, Atominstitut, TU Wien, Stadionallee 2, 1020 Vienna, Austria}

\author{RuGway Wu}
\affiliation{Vienna Center for Quantum Science and Technology, Atominstitut, TU Wien, Stadionallee 2, 1020 Vienna, Austria}

\author{J\"{o}rg Schmiedmayer}
\affiliation{Vienna Center for Quantum Science and Technology, Atominstitut, TU Wien, Stadionallee 2, 1020 Vienna, Austria}

\date{\today}

\begin{abstract}
We implement two types of matter-wave interferometers using trapped Bose-condensed Feshbach molecules, from weak to strong interactions. In each case, we focus on investigating interaction effects and their implications for the performance. In the Ramsey-type interferometer where interference between the two motional quantum states in an optical lattice is observed, interparticle interactions are found to induce energy shifts in the states. Consequently, this results in a reduction of the interferometer frequency and introduces a phase shift during the lattice pulses used for state manipulation. Furthermore, nonuniformity leads to dephasing and collisional effects contribute to the degradation of contrast. In the Michelson-type interferometer, where matter waves are spatially split and recombined in a waveguide, interference is observed in the presence of significant interaction, however coherence degrades with increasing interaction strength. Notably, coherence is also observed in thermal clouds, indicating the white-light nature of the implemented Michelson-type interferometer. 
\end{abstract}

\maketitle

\section{Introduction}
\label{sec:intro}

Since the first demonstration of Bose-Einstein condensate (BEC) interference~\cite{andrews_observation_1997}, BECs have emerged as an important source for interferometry~\cite{Cronin2009,Hogan2009book}, and many matter-wave interferometers using BECs have been implemented in a large variety of configurations. Here, we focus on interference with trapped atoms in different \textit{motional} states.  Examples include interference in waveguides~\cite{Wang2005, Garcia2006, Burke2008, Sapiro2009, McDonald2013, Li2014}, in double wells~\cite{Shin2004, AGF05, Schumm2005b, Esteve2008a, Gross2012, Berrada2013}, multiple transverse motional states~\cite{van_frank_interferometry_2014,Hu2018}, and Talbot effect~\cite{Deng1999,Mark2011,Yue2013,Wei2024}. Unlike optical interferometers, trapped matter waves experience particle-particle interactions that, on one hand result in phase shifts, limiting the system's coherence time~\cite{Dimopoulos2028,Grond2009MZ,Grond2010e,Jannin2015}, but on the other hand allow for the probing and characterization of many-body systems and their dynamics ~\cite{Hadzibabic2006, Hofferberth2007, Hofferberth2008, Langen2015a, Schweigler2017}. By suppressing the strength of interaction, long coherence times have been achieved ~\cite{Fattori2008, Gustavsson2008}. In addition, these interactions induce quantum properties such as squeezing and entanglement, beneficial for precision metrology~\cite{Esteve2008a, Gross2012, Berrada2013}. Controlling the splitting process, squeezing and entanglement can be enhanced~\cite{Grond2009a, Zhang2023}. All these experiments are conducted with weakly interacting systems where the scattering length is small compared to the unitary limit. In our present work, we study interference in systems of strongly interacting particles with large scattering lengths. $^6$Li$_2$ Feshbach molecules are particularly well suited for conducting experimental studies with bosons under strong interaction and observing dynamics, relaxation processes, and interference over extended periods. The strong suppression of inelastic collisions due to the fermionic nature of the constituent atoms~\cite{Petrov2004,Petrov2005} enables having long lifetimes of the samples. 

In the experiments presented here, we implement and study two distinct matter-wave interferometer schemes: Ramsey-type and Michelson-type interferometers, with strongly interacting $^6$Li$_2$ Feshbach molecules. We particularly focus on the effects of interparticle interactions on interferometer performance---frequency, phase shift, read-out contrast, and coherence. The measurements are compared with numerical simulations and calculations based on the experimental parameters. In addition, we observe interference and coherence in the Michelson-type interferometer scheme with noncondensed, thermal Feshbach molecules. 

Both interferometer schemes are implemented in our $^6$Li experimental apparatus, described with full details in Ref.~\cite{Liang2022}. The configurations in both schemes are highly similar. 
In both experiments, we start with a BEC of $^6$Li$_2$ Feshbach molecules. Each Feshbach molecule is constituted by two lithium atoms in the lowest hyperfine states $\ket{I=1,\,m_I=1,\,S=1/2,\,m_s=-1/2 }$ and $\ket{I=1,\,m_I=0,\,S=1/2,\,m_s=-1/2 }$. 
We set the interparticle interaction strength over a range by tuning the $s$-wave scattering length via the Feshbach resonance~\cite{Bartenstein2005}. Close to the Feshbach resonance, the intermolecular scattering length is approximately 0.6 times the scattering length between the two constituent atoms in different spin states~\cite{Petrov2004}. 
The molecular condensate is confined in a hybrid trap formed by an optical dipole potential from a red-detuned, focused laser beam aligned horizontally, and a horizontally confining magnetic potential arising from field curvature. Along the axial direction of the trap (the propagation direction of the dipole beam), the potential is primarily provided by magnetic field curvature and has a trap frequency $\omega_z \simeq 2\pi\times$18.5~Hz. Radial confinement is dominated by the dipole potential and has a radial trap frequency $\omega_\perp \simeq 2\pi\times$80~Hz. 
The interferometer operations are realized using an optical lattice formed by a pair of 1064-nm laser beams crossing at 15$^\circ$, resulting in $d=4.1~\mathrm{\mu m}$ lattice spacing. The lattice can be pulsed for state preparation, as well as maintained at a constant intensity to study the evolution of the superposition state of the Ramsey interferometer in the lattice~\cite{Denschlag2002}. We have verified that the scattering of the mBEC by lattice pulses remains mostly coherent and the molecules can be returned to the $k=0$ condensate~\cite{Liang2022}. This observation indicates that the molecules mostly remain bound throughout the process.

\section{Ramsey-Type Interferometer}
\label{sec:ramsey}
\subsection{Experimental implementation}
\label{sec:ramsey_experiment}

\begin{figure}[!tb]
\center
\includegraphics[width=1\columnwidth]{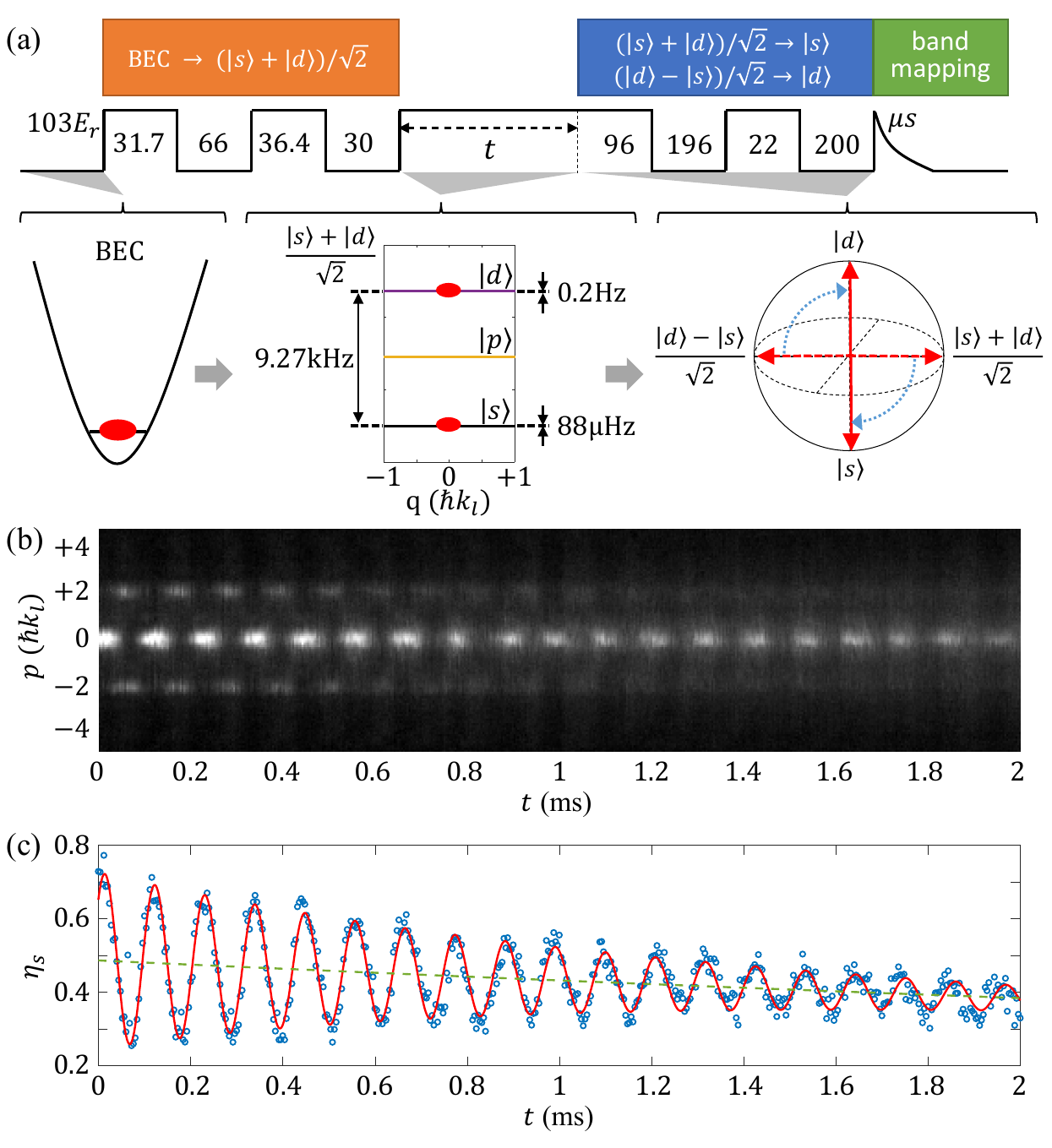}
\caption{(a) Experimental configuration of the Ramsey interferometer. Initially confined in a harmonic trap, a BEC is transferred to a superposition of the $|s\rangle$ and $|d\rangle$ states within an optical lattice using two lattice pulses. Treated as a two-level quantum system, the superposition state resides on the equator of a Bloch sphere, evolving at a frequency determined by the energy gap between the two motional states. Subsequently, a $\pi/2$ pulse, composed of two lattice pulses, is executed to bring the superposition state to the readout states. Finally, band mapping is utilized for detecting occupations in the two states. 
(b) Coherent oscillation in a Ramsey interferometer ($a=487a_0$, $N=7500$). Band-mapping distributions, revealing interference, are plotted against interference time $t$.
(c) Interference fringes shown with $|s\rangle$ state occupation $\eta_s$. Experimental results, depicted by the blue open circles, are fitted with a damped oscillation, represented by the red solid curve. The green dashed line indicates the decay of the oscillation center. 
}
\label{fig:ramsey_seq}
\end{figure}

In the Ramsey interferometer scheme, two motional states within the optical lattice are utilized: the ground state $|s\rangle$ and the second excited state $|d\rangle$, conceptualized as a spin-1/2 system~\cite{Hu2018}. 
The lattice depth, calibrated through Kapitza-Dirac (KD) scattering, is set at $V_{latt}=103E_r$. Here, $E_r=\hbar^2k_l^2/2m$ denotes the characteristic energy of the lattice, with $k_l=\pi/d$ the wave vector of the optical lattice and $m$ the molecular mass. 
In Fig.~\ref{fig:ramsey_seq}(a), we depict the Bloch band structure of our lattice configuration under zero interaction strength. The bandwidths of $|s\rangle$ and $|d\rangle$ states are significantly narrower compared to the energy gap separating them. Therefore, the broadening of the band transition due to the quasi-momentum distribution is not considered further.  

Initially, a nearly pure condensate with negligible thermal component is prepared, comprising 3500-9500 Feshbach molecules. The variation in particle numbers results from different loss rates, influenced by interparticle interaction. 
Using a shortcut loading method~\cite{Zhou2018}, as shown in Fig.~\ref{fig:ramsey_seq}(a), the BEC in the ground state of a weak harmonic confinement is transformed into the superposition state $|\psi\rangle=(|s\rangle+|d\rangle)/\sqrt{2}$ within the optical lattice using two lattice pulses. Instead of first loading the BEC into the $|s\rangle$ state and then sequentially executing a $\pi /2$ rotation, this combined operation is found to be more robust and efficient. 
The lattice pulse sequence, tailored for maximum operation fidelity under noninteracting conditions, can achieve up to 98.4\% fidelity based on calculations. 
For further details on the calculations of pulse sequence, see our previous work~\cite{Liang2022}. 
The superposition state is maintained in the optical lattice for a variable interference time $t$. 
A subsequent two-pulse $\pi/2$ operation then transfers the state $(|s\rangle+|d\rangle)/\sqrt{2}$ to $|s\rangle$ and the $(-|s\rangle+|d\rangle)/\sqrt{2}$ to $|d\rangle$. 
Immediately afterward, the state occupation is read out using band mapping~\cite{Greiner2001}. 
After an adiabatic opening of the lattice and a time of flight (TOF), particles in different states separate and are detected via standard absorption imaging. 

\subsection{Interaction effects in Ramsey interferometer}
\label{sec:ramsey_example}

We first perform the Ramsey interferometer experiment with a molecular number $N=7500$ and an inter-molecular scattering length $a=487a_0$. Here, $a_0$ denotes the Bohr radius. 
Unless specified otherwise, $a$ refers to the $s$-wave scattering length between molecules. 
Figure~\ref{fig:ramsey_seq}(b) displays a time carpet showing the oscillating occupations of the $|s\rangle$ and $|d\rangle$ states over the interference time, which evidences the Ramsey interference.
Particles in the $|s\rangle$ state occupy the central zone (the first Brillouin zone), while those in the $|d\rangle$ state appear around $p=\pm2\hbar k_l$. 
The proportion of particles in the $|s\rangle$ state, denoted as $\eta_s(t)$, is determined by the ratio $N_s(t)/N(t)$, where $N_s(t)$ is the number of particles found in the first Brillouin zone and $N(t)$ is the total particle number at that time. Figure~\ref{fig:ramsey_seq}(c) shows a damped oscillation of $\eta_s$ as a function of the interference time $t$, with its fitting curve by the function 
\begin{equation}
\begin{aligned}
\eta_s = A e^{-\lambda_a t} \cos(\omega t+\phi)/2
       +B e^{-\lambda_b t}+C\,.
\label{eq:fit}
\end{aligned}
\end{equation}
In the analysis of our experimental data, we focus on the following key parameters extracted from fitting the observed interference patterns: 

(1) Interference frequency $\omega$: The interference frequency is determined to be $\omega=2\pi \times 9.096(33)$~kHz. Compared to the case of zero interaction where the energy gap is calculated to be $\omega_0=2\pi \times 9.27$~kHz, the $|s\rangle$ and $|d\rangle$ states are subject to different energy shifts due to interactions. Because of the broader localized wave functions in higher states, the energy level of the $|s\rangle$ state shifts upward to a greater extent than the $|d\rangle$ state, resulting in a reduced energy gap under stronger interactions. 
By treating the interactions as mean fields, the calculated energy gap shifts to $2\pi \times$9.118~kHz, which is in good agreement with our experimental measurement within the error margins. 
The differential energy shift between the two states can also be interpreted as an effective reduction of the lattice depth due to the interactions counteracting the lattice potential, which has been demonstrated in our previous work by observing the slowing-down effect in the time evolution of the momentum mode populations under strong interactions~\cite{Liang2022}. The variation of interference frequency relative to interactions will be further explored in Secs.~\ref{sec:ramsey_interaction} and \ref{sec:ramsey_calculation} and summarized in Fig.~\ref{fig:ramsey_aVar}. 

(2) Phase shift $\phi$: We observe the emergence of an additional phase shift, determined to be $\phi=-0.20(2)\pi$ based on the fit results. The lattice pulse sequences are designed to align the two $\pi/2$ operations (the first $\pi/2$ operation is implicit in the first two-pulse sequence) under the condition of vanishing interaction, setting the initial phase to zero. Nonetheless, interactions occurring during the lattice pulses can introduce a phase shift in the diffraction processes. The variation of phase shift relative to interactions will be discussed in Sec.~\ref{sec:ramsey_interaction} and summarized in Fig.~\ref{fig:ramsey_aVar2}(a). 

(3) Maximal contrast $A$: The maximum contrast of the interference pattern is obtained at $t=0$, determined to be 0.44(3). The loss of contrast at this point is attributed to collisional losses of condensed particles during the lattice pulses and the subsequent TOF, which in turn attenuates the readout signal. During the separation stage, molecules with differing speeds undergo elastic collisions, leading them to scatter in random directions. This process reduces the population of the coherent condensate. 
Moreover, imperfections in the $(|s\rangle+|d\rangle)/\sqrt{2}$ state preparation also contribute to the reduced contrast, especially at very strong interactions. These imperfections stem from the fact that the lattice pulse sequences for state manipulation are designed without considering interactions. As interaction strength increases, a minor imbalance between the $|s\rangle$ and $|d\rangle$ states, along with slight occupations on higher-momentum modes, are expected. While the imperfection of state preparation is not remarkable under the conditions presented in Fig.~\ref{fig:ramsey_seq}, it presents a certain significance at stronger interaction strengths. 
Additionally, during state manipulation pulse sequences, molecules transiently occupied the fourth ($|g\rangle$) and higher excited states, which are known to exhibit significantly shorter lifetimes compared to the $|s\rangle$ and $|d\rangle$ states used in the interferometer. Inelastic collisions by the higher-energy molecules during the process can also contribute to decoherence. 
The variation of maximal contrast relative to interactions will be discussed in Sec.~\ref{sec:ramsey_interaction} and summarized in Fig.~\ref{fig:ramsey_aVar2}(b). 

(4) Contrast decay rate $\lambda_a$: The interference contrast experiences an exponential decay over time, characterized by a contrast decay rate $\lambda_a$. This decay primarily arises from interactions: The inhomogeneous density distribution and the density fluctuation of particles across lattice sites result in variations of the interaction energy within the cloud, which leads to dephasing. Averaging across the entire cloud in the readout, this dephasing process leads to a gradual reduction in interference contrast. 
Furthermore, collisional effects also contribute to contrast decay, particularly pronounced under strong interactions. In addition to the collisional losses discussed in the previous paragraph, the relaxation of particles from the $|d\rangle$ state to the $|s\rangle$ state due to inelastic collisions also becomes more pronounced as interactions become stronger. The calculated lifetime of the $|d\rangle$ state is provided in the Appendix. 
Additionally, the radial expansion of the cloud following shortcut loading, although not specifically considered in the calculations in Sec.~\ref{sec:ramsey_calculation}, may have some impact on the decay of contrast. The variation of contrast decay rate relative to interactions will be discussed in Sec.~\ref{sec:ramsey_interaction} and summarized in Fig.~\ref{fig:ramsey_aVar2}(c). 
Besides the effects arising from interactions, other mechanisms contributing to contrast decay in weakly interacting (noninteracting) systems have been explored in Ref~\cite{Hu2018}, including intensity fluctuations of trapping lasers, thermal fluctuations, and quantum fluctuations. Compared to the effects arising from interactions, these mechanisms make only minor contributions.

(5) Parameters describing the monotonically decreasing offset: $\lambda_b$, $B$, and $C$. 
The offset occurs because the number of molecules scattered into a broad incoherent background increases with time. The scattered molecules overlap more with the higher states than with the $|s\rangle$ state after TOF, leading to a downward shift of the $|s\rangle$ state population curve. 
The green dashed curve in Fig.~\ref{fig:ramsey_seq}(c) depicts this offset decay, mathematically represented by the expression $B  e^{-\lambda_b t} + C$.

\subsection{Ramsey interference at varying interaction strengths}
\label{sec:ramsey_interaction}

\begin{figure}[!tb]
\center
\includegraphics[width=1\columnwidth]{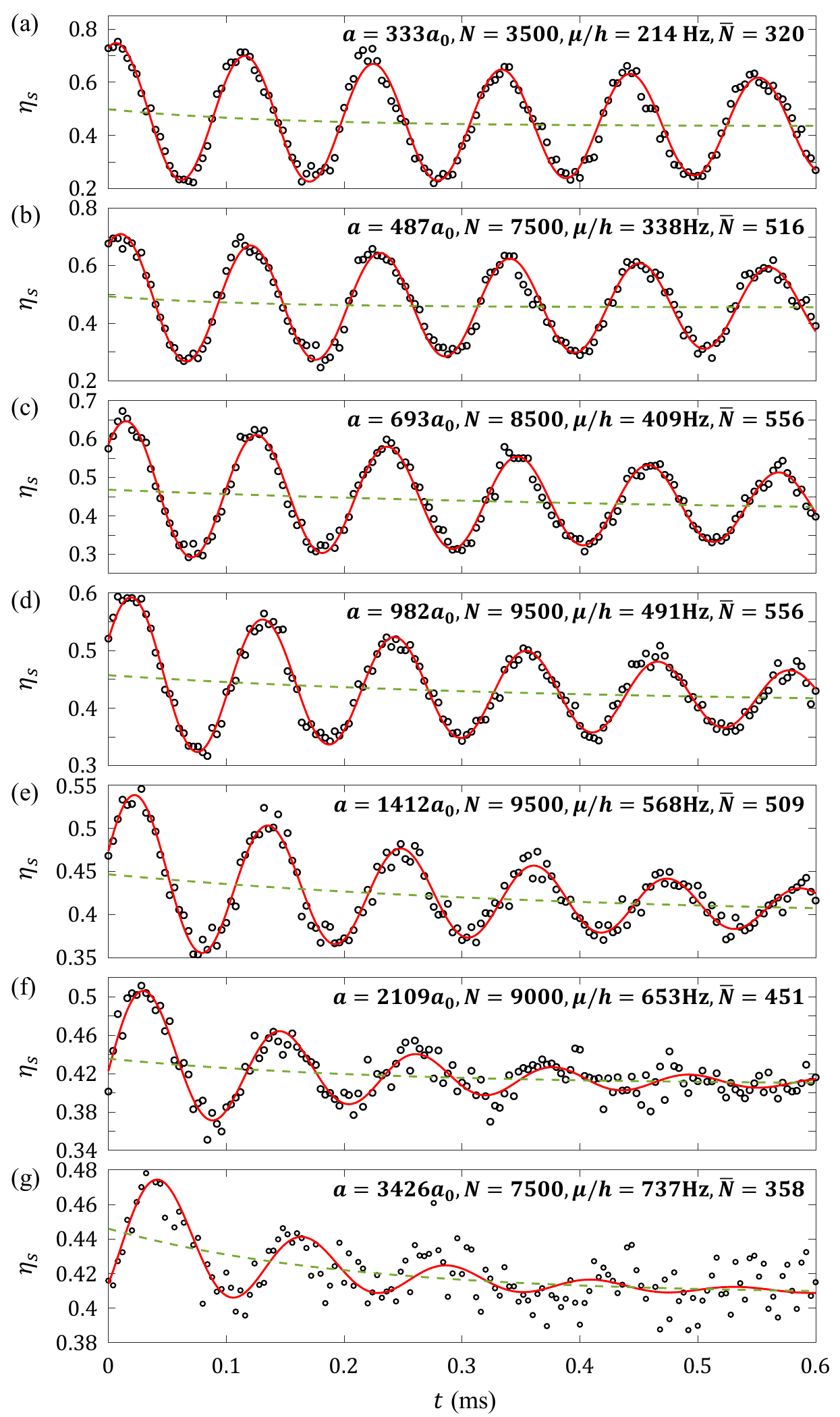}
\caption{
Coherent oscillations in Ramsey interferometers at varying interaction strengths. (a-g) Results spanning from weak to strong interaction strengths. The scattering length $a$, total particle number $N$, chemical potential $\mu$ prior to lattice loading, and weighted-average particle number per lattice site $\bar N$ are provided for each dataset. $h$ is the Planck constant. 
Interference frequencies, phase shifts, and decaying contrasts are extracted from the fit curves of damped oscillations. }
\label{fig:ramsey_osc}
\end{figure}

\begin{figure}[!tb]
\center
\includegraphics[width=0.6\columnwidth]{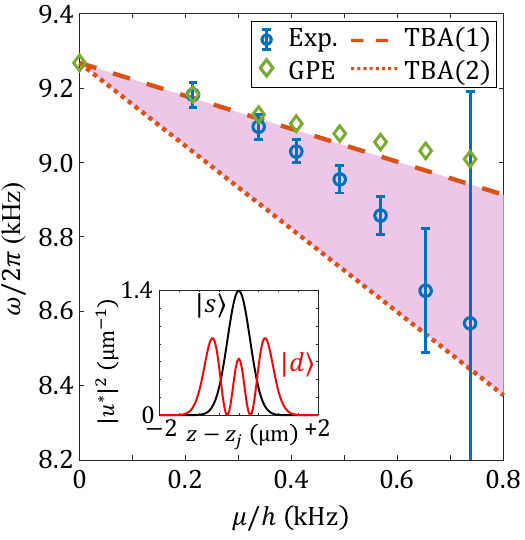}
\caption{
The Ramsey interference frequency varies with interaction strength. We plot the interference frequency against the chemical potential of the molecular condensate prior to lattice loading. The experimental results, represented by open circles, are extracted from Fig.~\ref{fig:ramsey_osc}, with error bars indicating a 95\% confidence level in fittings. The results fall within the shaded area between two lines obtained by calculations of the energy shift due to interaction under the tight-binding approximation (TBA). The upper line takes into account the interaction shifts of the $|s\rangle$ and $|d\rangle$ states, while the lower line includes only the energy shift of the $|s\rangle$ state. Due to more rapid decay of the $|d\rangle$ state population at higher interaction strength, the experimental results for higher chemical potential approach the lower line. A 1D Gross-Pitaevskii equation (GPE) calculation aligns with the upper line. The inset shows the localized wave functions of the $|s\rangle$ and $|d\rangle$ states in the axial direction. 
}
\label{fig:ramsey_aVar}
\end{figure}

\begin{figure}[!tb]
\center
\includegraphics[width=1\columnwidth]{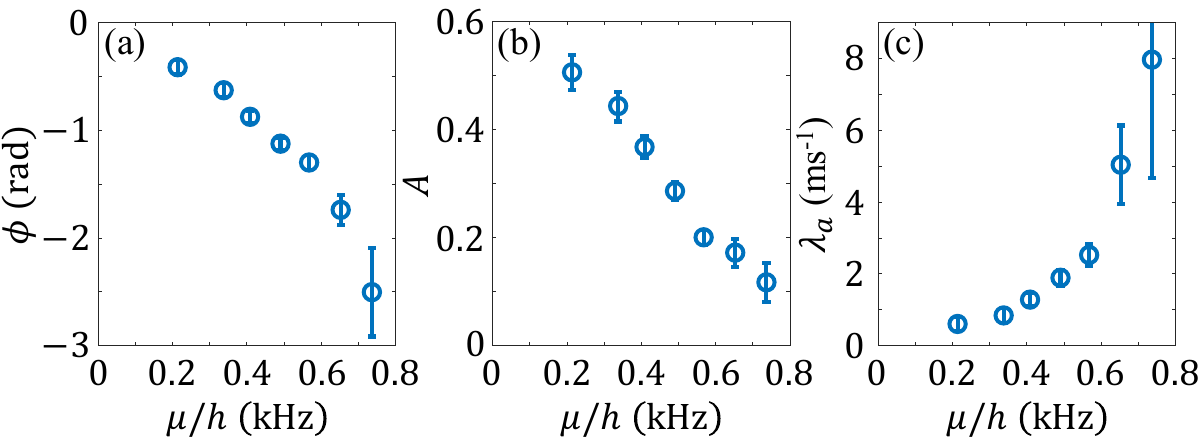}
\caption{
Interaction effects in the Ramsey interferometer. The plots illustrate (a) phase shift, (b) maximal contrast, and (c) contrast decay rate as functions of the chemical potential of the molecular condensate prior to lattice loading. Experimental results, depicted by open circles, are extracted from Fig.~\ref{fig:ramsey_osc}, with error bars indicating a 95\% confidence in fittings. 
}
\label{fig:ramsey_aVar2}
\end{figure}

To further explore the interaction effects, we perform additional Ramsey interferometric measurements over a range of scattering lengths from $330a_0$ to $3426a_0$, by adjusting the magnetic field near a Feshbach resonance. 
The measured interference signals are shown in Fig.~\ref{fig:ramsey_osc}.  
It is important to note that the initial conditions differ across data sets due to changes in the BEC ground-state spatial profiles and particle loss rates with varying interactions. For each data set, the interparticle scattering lengths $a$, the total molecule numbers $N$, the calculated chemical potential $\mu$ of the molecular condensate prior to lattice loading, and the weighted-average number per lattice site $\bar N$ ($\bar N = \sum_j N_j^2 / \sum_j N_j$, where $j$ is the site index) are provided in the figure. 
Continuing with the methodology outlined in Sec.~\ref{sec:ramsey_example}, we fit the experimental data using damped oscillations, shown as red solid curves in Fig.~\ref{fig:ramsey_osc}. From these fits, parameters such as $\omega$, $\phi$, $\lambda_a$, and $A$ are determined. 
In this study, we primarily focus on examining how the interference frequency $\omega$ varies in response to changes in the interparticle interaction strengths and comparing our measurement results with theoretical calculations. As for other parameters, we limit ourselves to presenting measurement results without conducting related calculations. This is due to the difficulty of modeling collisional losses. 
In the following discussion of parameters, the paragraph numbering for corresponding parameters continues from Sec.~\ref{sec:ramsey_example}. 

(1) The interference frequency $\omega$ is found to decrease with increasing interaction strength. In Fig.~\ref{fig:ramsey_aVar}, the measurements are presented against the calculated chemical potential $\mu$ of the molecular condensate prior to loading into the lattice. The observed frequency changes are minimally affected by collisional losses, directly reflecting the shifts in energy levels within the optical lattice induced by interactions. The measurements fall within the shaded area, indicating the expected interference frequencies between two calculations of energy shift due to interaction. The calculations are detailed in Sec.~\ref{sec:ramsey_calculation}.

(2) The phase shift $\phi$ becomes more pronounced as the interaction strength increases, as shown in Fig.~\ref{fig:ramsey_aVar2}(a). The sequential pulses, designed to result in zero differential phase in the absence of interaction, introduce additional phase shifts due to energy shifts at finite interaction strengths. This observation is consistent with previous findings~\cite{Buechner2003, Yue2013}. When extrapolating the curve along its trend, we observe an intercept close to zero, aligning with the intended behavior of the pulse sequence designed for zero interactions. 

(3) The maximum contrast $A$ is observed to decrease as the interaction strength increases, as shown in Fig.~\ref{fig:ramsey_aVar2}(b). This reduction is mainly attributed to the increased collisional losses and imperfections in the initial state under strong interactions. Interference measurements are significantly constrained under strong interactions, as particles initially occupying the condensate peaks are scattered into the background, resulting in a reduction of the signal-to-noise ratio. 

(4) The contrast decay rates $\lambda_a$ are enhanced under strong interactions for the reasons stated in Sec.~\ref{sec:ramsey_example}. The dephasing and collisional effects play much more crucial roles in the contrast decay under strong interactions, compared to the mechanisms independent of interactions. Based on the observed trend of the data points shown in Fig.~\ref{fig:ramsey_aVar2}(c), we anticipate minor contrast decay at zero interaction strength.

\subsection{Numerical calculations}
\label{sec:ramsey_calculation}

Using the Bloch theorem~\cite{Dalibard2013}, the single-particle spectrum and eigenfunctions in a one-dimensional (1D) optical lattice can be simply calculated. The energy band structure at a lattice depth of $V_{\text{latt}} = 103E_r$ with an interaction strength of zero is illustrated in Fig.~\ref{fig:ramsey_seq}(a). Here, the energy gap between the $|s\rangle$ and $|d\rangle$ states amounts to $\omega_0 = 2\pi \times 9.27~\text{kHz}$. The Ramsey frequency is given by the energy gap between the two states. To determine the frequency shift incurred by interaction, we calculate the differential energy shift of the two states induced by interaction. 

In the deep lattice limit, a condensate in a 1D optical lattice can be described by the 1D discrete nonlinear Schr\"{o}dinger equation (DNLS)~\cite{Trombettoni2001}. Under the tight-binding approximation (TBA)~\cite{Smerzi2003}, the system behaviors are described by the localized wave function in each lattice site,
\begin{equation}
\Psi(\boldsymbol{r},t)=\sum_j \psi_j(t) \Phi_j(\boldsymbol{r}) = \sum_j\sqrt{N_j(t)}e^{i \varphi_j(t)}\Phi_j(\boldsymbol{r})\,,
\end{equation}
where $N_j$ and $\varphi_j$ denote the particle number and the phase at the $j^\mathrm{th}$ lattice site. The spatial wave function,   
\begin{equation}
\Phi_j(\boldsymbol{r})\backsimeq u^*(z)  \phi_j(\boldsymbol{r_\perp})\,,
\end{equation}
is factorized into a product of the axial function $u^*(z)$ and the radial function $\phi_j(\boldsymbol{r_\perp})$. $u^*(z)$ is obtained from the Bloch wave function $e^{ik z} u(z)$: it represents one period of the periodic term $u(z)$ and is localized within a single lattice site ($-d/2<z-z_j<d/2$, where $z_j$ denotes the trap bottom of the $j^\mathrm{th}$ lattice site). $|u^*(z)|^2$ is displayed in the inset of Fig.~\ref{fig:ramsey_aVar}, and $\int |u^*(z)|^2 dz$ is normalized to unity. In the deep lattice regime we are working with, $u^*(z)$ approaches zero at the lattice barriers, validating the tight-binding approximation in our calculations. Additionally, $u^*(z)$ remains independent of the particle number, interaction strength and, quasimomentum in our deep optical lattice. 

Given that the lattice loading time is significantly shorter than the time scale associated with the radial trap frequency, we take $\phi_j(\boldsymbol{r_\perp})$ to be the cross-sectional profile at $z=z_j$ of the initial condensate wave function prior to lattice loading. This profile is determined by the prelattice potential geometry, the total particle number $N$, and the scattering strength $a$. 
By treating the interactions as mean fields, the interaction chemical potential at the $j^\mathrm{th}$ lattice site is calculated as 
\begin{equation}
\mu_{j}^{int}=|\psi_j|^2 g_0 \int|\Phi_{j}(\boldsymbol{r})|^4 d \boldsymbol{r}\,, 
\label{eq:mu}
\end{equation}
and it determines the energy shift of the state due to interaction. Here, $g_0=4\pi \hbar^2 a/m$. 

We now extend this calculation framework to determine the differential energy shift between the $|s\rangle$ and $|d\rangle$ states. Due to the finite lifetime of particles in the $|d\rangle$ state (as provided in the Appendix), we examine two boundary cases: (1) We consider the energy shifts of both the $|s\rangle$ and $|d\rangle$ states: $\Delta E^{int}_j=\mu^s_j-\mu^d_j$, with $\mu^{s(d)}_j$ calculated from Eq.~(\ref{eq:mu}). The interaction between the two states shifts them by the same amount, thereby leaving their energy gap unchanged. The particle numbers in both states are $N_j/2$, and their radial wavefunctions both remain $\phi_j(\boldsymbol{r}_\perp)$. (2) We only account for the energy shift of the $|s\rangle$ state, $\Delta E^{int}_j=\mu^s_j$, while neglecting the $|d\rangle$ state due to particle loss and its overexpanded radial profile resulting from collisions. In both cases, we evaluate this interaction effect throughout the entire cloud, by calculating the weighted-average differential interaction energy as
\begin{equation}
\Delta E^{int}=\frac{\sum_j N_j\Delta E^{int}_j}{\sum_j N_j}\,. 
\end{equation}

In Fig.~\ref{fig:ramsey_aVar}, we observe that the experimental measurements fall between the two calculated results. For the data point with the weakest interaction strength we study ($\mu/h=214$Hz), the measurement aligns perfectly with the calculation considering the energy shifts of both states [TBA(1)]. For this measurement the lifetime of particles in the $|d\rangle$ state is calculated to be 21ms, much longer than the experimental time scale. Therefore, considering the two states with identical particle number and radial profile is a good approximation.  As the interaction strength increases, the measurements gradually approach the other boundary [TBA(2)]. For the data point with the strongest interaction strength, we study ($\mu/h=737$Hz), the lifetime of particles in the $|d\rangle$ state is reduced to 0.6ms, matching the observation time scale. The particle number in the $|d\rangle$ state decreases significantly, and its radial profile expands due to collisions. Consequently, the resulting measurement is closer to the calculation considering only the energy shift of the $|s\rangle$ state. These calculations effectively capture the interaction effects on the interference frequency shift and the collisional decay. 

For comparison, we also perform mean-field calculations using continuous wave functions based on the 1D Gross-Pitaevskii equation (GPE). Starting with a BEC in the ground state of a harmonic trap, the optical lattice is implemented following the experimental protocol. Band occupations are measured by ramping down the lattice, and the interference frequency is analyzed in the approach with the experimental data. The results, presented as diamonds in Fig.~\ref{fig:ramsey_aVar}, are consistent with the upper line of the energy shift calculations under the tight-binding approximation. This alignment occurs because the decay of particles in the $|d\rangle$ state is not prominent and the dynamics in the radial directions are constrained in 1D GPE calculations. 
\section{Michelson-Type Interferometer}
\label{sec:mz}
\subsection{Experimental implementation}
\label{sec:MZ_experiment}

A Michelson-type interferometer in a waveguide is also implemented, investigating the interference phenomena occurring between spatially separated paths. The splitting and recombining of a molecular condensate are achieved by Kapitza-Dirac (KD) scattering pulses. During the process between splitting and recombination, the two emerging clouds undergo half an oscillation in the weak harmonic potential along the waveguide. A differential phase shift between two separated arms of the interferometer is introduced by a magnetic field gradient pulse.  
The operation adheres to the procedure outlined in pioneering works~\cite{Wang2005, Garcia2006, Burke2008, Sapiro2009, McDonald2013, Li2014}, with a key distinction in our interferometer: the matter waves consist of Feshbach molecules, exhibiting significant interactions. 
An additional feature of the Michelson-type interferometer in comparison to the Ramsey interferometer is the relative bulk motion of the wave packets. This strong motion leads to a more pronounced loss of condensed molecules due to collisions, as the scattering rate scales with the relative velocity of the colliding molecules, as well as with the number density and the collisional cross section. With the Michelson-type scheme, one has the opportunity to look into the collisional loss during the passage of the wave packets through each other, and its effect on coherence. 

\begin{figure}[!tb]
\center
\includegraphics[width=1\columnwidth]{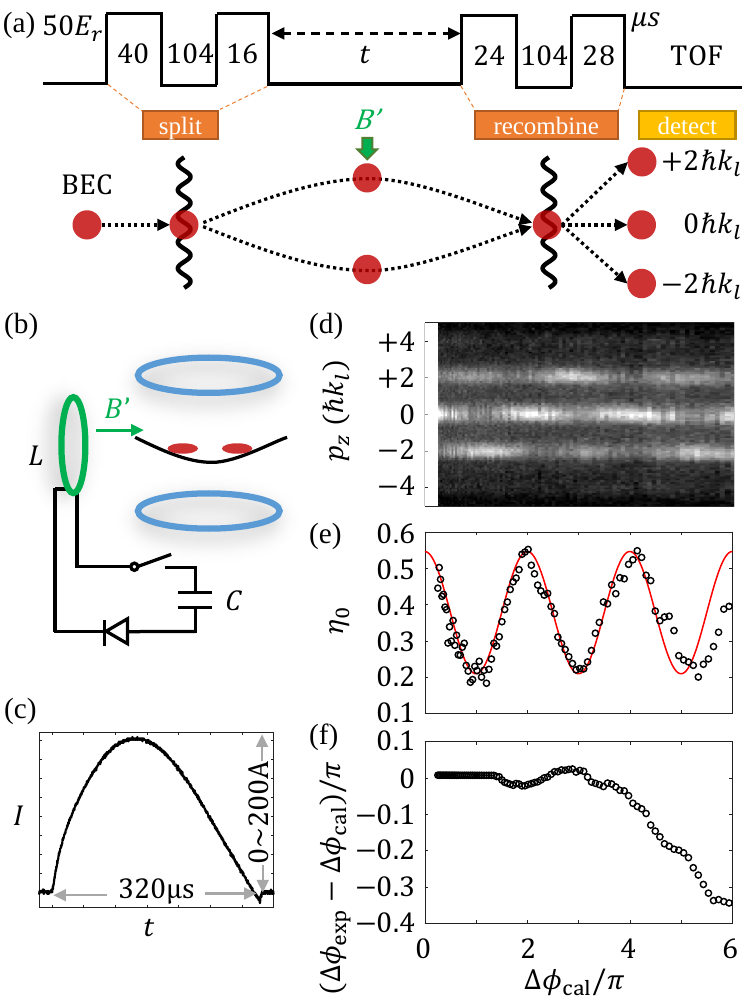}
\caption{
(a) Experimental configuration of the waveguide Michelson-type interferometer. A BEC is split into two wave packets by a pair of lattice pulses, undergoing half an oscillation in the axial confinement of a waveguide before recombination via another pair of lattice pulses. A differential phase shift between the two interferometer arms is introduced with a magnetic field gradient pulse at the maximum separation of the two wave packets. The interference patterns are analyzed by examining the occupations in momentum modes. 
(b) Coil setup. Currents flowing through a pair of coils (indicated by two blue circles) produce the Feshbach field for adjusting interparticle interactions. A slight deviation from the Helmholtz condition creates a harmonic potential along the waveguide axis. A gradient coil (indicated by a green circle), with its axis aligned with the waveguide, in conjunction with a precharged capacitor, constitutes an LC resonant circuit. This setup is used to apply a magnetic field gradient pulse, inducing a differential phase shift between the two interferometer arms. 
(c) An example of the current pulse flowing through the gradient coil. 
(d) Coherent oscillation in a Michelson-type interferometer ($a=487a_0$ and $N=6000$). Momentum distributions, revealing interference, are plotted against the applied differential phase shift. 
(e) Interference fringes are shown by the oscillating occupation of the $|0 \hbar k_l \rangle$ mode, denoted by $\eta_0$. The fringe produced with the calculated phase shift and the experimentally measured amplitude is plotted for comparison in a red solid curve. 
(f) Discrepancy between the experimentally measured phase shifts $\Delta \phi_\mathrm{exp}$ and the calculated phase shifts $\Delta \phi_\mathrm{cal}$. 
}
\label{fig_osc}
\end{figure}

The experimental setup employs an mBEC consisting of 6000 Feshbach molecules, with the scattering length between molecules ranging from $487a_0$ to $982a_0$. The processes of splitting, phase shifting, recombining, and detecting the matter waves are illustrated in Fig.~\ref{fig_osc}(a). The interference occurs within a waveguide potential given by $U(r_\perp,z) = m(\omega_\perp^2r_\perp^2 + \omega_z^2z^2)/2$. Initially at rest, the condensed molecules are exposed to two sequential KD scattering pulses. The lattice depth during this process is maintained at $50E_r$. This process, following the Newton's cradle experiment~\cite{Li2020}, splits the molecules into two momentum modes $|\pm 2\hbar k_l\rangle$ along the waveguide. After the splitting, the two matter-wave packets separate along the waveguide and ascend the axial harmonic potential. 

To demonstrate interference and controllability, a differential phase shift is introduced between the two arms of the interferometer when the two matter-wave packets are at their maximum separation of $130~\mathrm{\mu m}$. This phase shift is induced by a magnetic field gradient pulse, generated by an electrical current pulse flowing through a gradient coil, whose central axis is aligned with the interferometer arms. The coil setup used in our Michelson-type interferometer experiment is shown in Fig.~\ref{fig_osc}(b). The gradient coil, in conjunction with a precharged capacitor, forms an LC resonant circuit. A diode is integrated into the circuit in series to ensure unidirectional current flow. Each time the circuit is closed, it produces a pulse of a field gradient along the waveguide that lasts for half of an LC oscillation cycle, as shown in Fig.~\ref{fig_osc}(c). The duration of the pulse, determined by the capacitance and inductance, remains constant throughout the experiment. The pulse strength, which is adjusted by varying the charging voltage applied to the capacitor, determines the differential phase shift. 
The separation between the two packets is considerably larger than their individual sizes, and the duration of the field gradient pulse is significantly shorter than the propagation time between the splitting and recombining of the matter-wave packets. Therefore, we make the assumption in the calculation that the two wave packets remain stationary during the pulse. 
The differential phase shift is given by 
\begin{equation}
    \Delta \phi =\int_t \frac{\mu_{Li_2}[B_1(t)-B_2(t)]}{\hbar}dt\,.
\label{eq_phase}
\end{equation}
Here, $\mu_{Li_2}$ is the magnetic dipole moment of the $^6$Li$_2$ molecule, and $B_1$ and $B_2$ are the time-dependent magnetic fields at the two separated locations of the matter-wave packets. 

Following the application of the phase shift, the two wave packets reverse their directions of motion and propagate back toward their origin. Upon reaching the origin, the two wave packets are recombined using a second pair of KD pulses. 

The interference pattern is analyzed by examining the distribution of particles across various momentum modes as a function of the differential phase shift. Following the recombination pulses, all optical potentials are abruptly turned off, allowing the cloud to expand while the magnetic field is maintained. During the TOF, the field curvature provides a focusing potential in the horizontal plane for momentum distribution measurement~\cite{Murthy2014}. Consequently, the cloud quickly expands in the radial direction, resulting in a rapid decrease in interparticle interactions. After a quarter of the oscillation period, determined by the horizontal trapping frequency of 18.5 Hz, the spatial profile of the cloud along the axial direction reflects the momentum at the moment when the optical potentials are turned off. Detection is achieved by employing absorption imaging. 

\subsection{Michelson-type interference at varying interaction strengths}
\label{sec:MZ_interaction}

We first demonstrate the interferometric results measured at $a=487a_0$. 
Figure~\ref{fig_osc}(d) displays the oscillating occupations of the $| 0 \rangle$ and $|\pm 2\hbar k_l \rangle$ modes as the differential phase shift increases (via an increase in the magnetic field gradient strength). 
In addition, a very small occupation in the $|\pm 4\hbar k_l \rangle$ modes is noted, attributed to the imperfect design of the splitting and recombining processes. 
The imbalance between $|- 2\hbar k_l \rangle$ and $|+ 2\hbar k_l \rangle$ arises from the disturbance in the motion of the cloud caused by the magnetic field gradient. 

To determine the occupation in each momentum mode, we fit the momentum distribution at each time point using a profile comprising five narrow condensate peaks positioned at $p=-4 \hbar k_l,\,-2\hbar k_l,\,0\hbar k_l,\,+2\hbar k_l,\,+4\hbar k_l$ (with particle numbers $N_{-2}$, $N_{-1}$, $N_0$, $N_1$, $N_2$) superimposed on a broad Gaussian peak that represents the scattered particles. We exclude the scattered particles, and only account for the particles in the condensate modes to calculate the fractional occupation of each state. For instance, the population in the $|0\hbar k_l\rangle$ mode is determined as $\eta_0 = N_0/\sum_{m=-2}^{+2}N_m$ [illustrated in Fig.~\ref{fig_osc}(e)]. 
The results in Figs.~\ref{fig_osc}(d)-\ref{fig_osc}(f) are shown as a function of the calculated differential phase shift between the two interferometer arms $\Delta\phi_\mathrm{cal}$. 
The phase shift calculation is conducted through a numerical integration based on the Biot-Savart law along our coil wire to determine the magnetic fields at the locations of the two wave packets. This process takes into account the precise positioning and dimensions of each coil turn. 
In Fig.~\ref{fig_osc}(e), a reference interference signal, produced from the calculated phase shift and the experimentally measured amplitude, is presented alongside the measurements for visual comparison.  
In Fig.~\ref{fig_osc}(f), we present the discrepancy between the experimentally measured phase shifts $\Delta \phi_\mathrm{exp}$ and the calculated phase shifts $\Delta \phi_\mathrm{cal}$. 
The measured phase shift $\Delta \phi_\mathrm{exp}$ is determined by fitting a section of experimental data within a close range ($\pm 3\pi /2$) to a cosine curve. When $\Delta \phi <3\pi /2$, the fitting section remains fixed from 0 to $3\pi$, resulting in a straight line at the beginning of the result. As $\Delta \phi$ increases beyond $3\pi /2$, the fitting section gradually shifts, scanning over all experimental measurements step by step.

In the measurements presented in Figs.~\ref{fig_osc}(d)-\ref{fig_osc}(f), we note that for phase shifts $\Delta \phi<3.5\pi$, the calculations closely match the observations with an error margin of $\pm 0.02 \pi$. However, for larger phase shifts, where stronger magnetic field gradients are applied, the actual phase shift increases at a slower rate than calculated. This growing discrepancy is attributed to the disturbances to the wave-packet trajectories caused by strong magnetic field gradient pulses. 

\begin{figure}[!tb]
\center
\includegraphics[width=1\columnwidth]{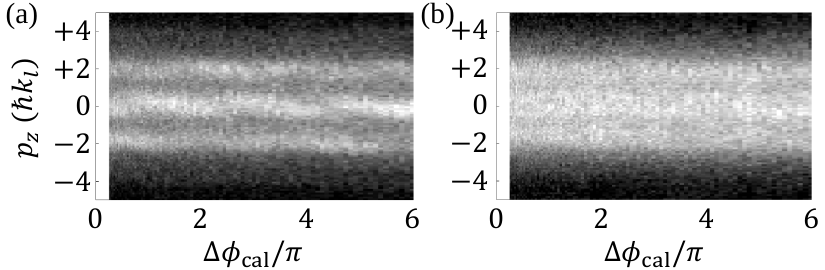}
\caption{
Interferometric results for the Michelson-type configuration at stronger interaction strengths, specifically at scattering lengths (a) $a=693a_0$ and (b) $a=982a_0$. 
}
\label{fig_aVar}
\end{figure}

In Refs.~\cite{Burke2008} and \cite{Horikoshi2007}, it has been demonstrated that under weak interactions, the main limitations on the coherence time of guided-wave atom interferometry arise from phase gradients caused by axial confinement and interparticle interactions. The dephasing of wave packets due to phase gradients could be suppressed by ensuring sufficient separation between the two wave packets in space and allowing them to undergo symmetric motion in the waveguide, such as completing a full oscillation. 
However, as the interparticle interaction becomes stronger, collisional loss emerges as the predominant factor in decoherence processes. Since collisional loss is not dependent on position, it cannot be suppressed through a symmetric trajectory design. In our experiments, it is confirmed that applying the recombining pulses upon the completion of one full oscillation results in a further reduction in fringe contrast, dropping to as low as 20\%. This indicates a more pronounced decrease compared to the scenario of a half-period oscillation. This observation underscores the collisional loss as a key limiting factor in the performance of interferometry with strong interaction.

This reduction in contrast due to collisional losses becomes particularly pronounced as we continue to increase the interaction strength. In Fig.~\ref{fig_aVar}, we present additional measurements performed at stronger interactions, specifically at scattering lengths of $a=693a_0$ and $982a_0$. 
These results indicate that stronger interactions lead to increased condensate loss, significantly diminishing interference contrasts~\cite{Hoellmer2019}. 
Our observations highlight the challenges in achieving extended coherence times under strong interaction conditions.

\subsection{Interference in thermal clouds}
\label{sec:MZ_halfperiod}

\begin{figure*}[!tb]
\center
\includegraphics[width=1.7\columnwidth]{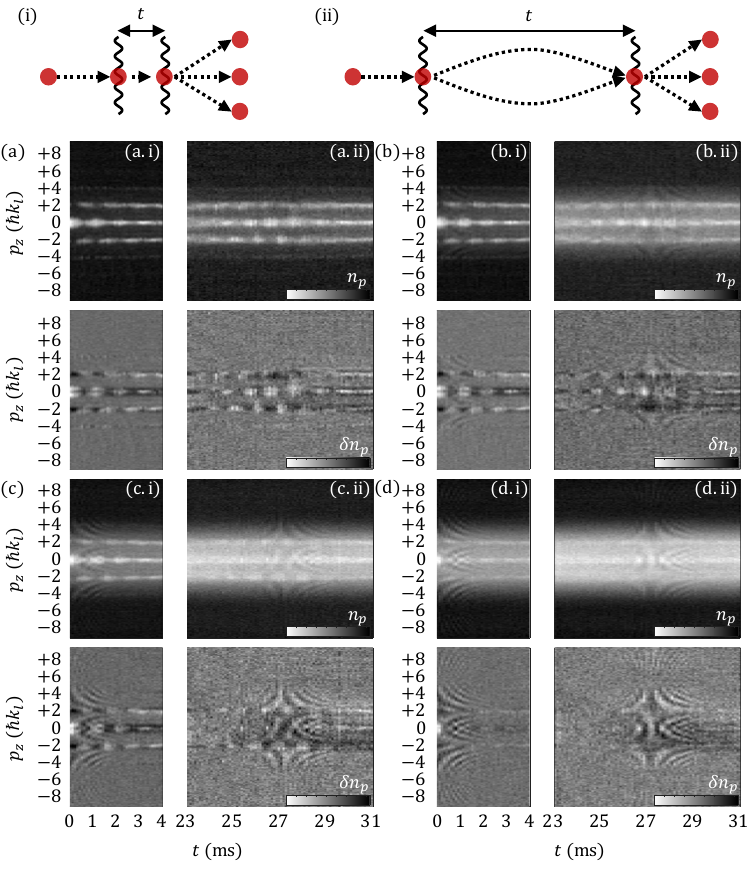}
\caption{
Interference patterns observed in Michelson-type interferometers, transitioning from condensates to thermal clouds, are shown in (a)-(d). The temperatures are estimated to be (a) 20nK, (b) 40nK, (c) 65nK, and (d) 90nK. Each data set captures interference at two propagation times: (i) shortly after splitting and (ii) around half an oscillation later. To enhance fringe visibility, we calculate $\delta n_p(p_z,t) = n_p(p_z,t) - \langle n_p(p_z,t)\rangle_t$, where $\langle \cdot \rangle_t$ denotes time average. The corresponding results are displayed below each absorption imaging carpet.
}
\label{fig_temp}
\end{figure*}

\begin{figure}[!tb]
\center
\includegraphics[width=1\columnwidth]{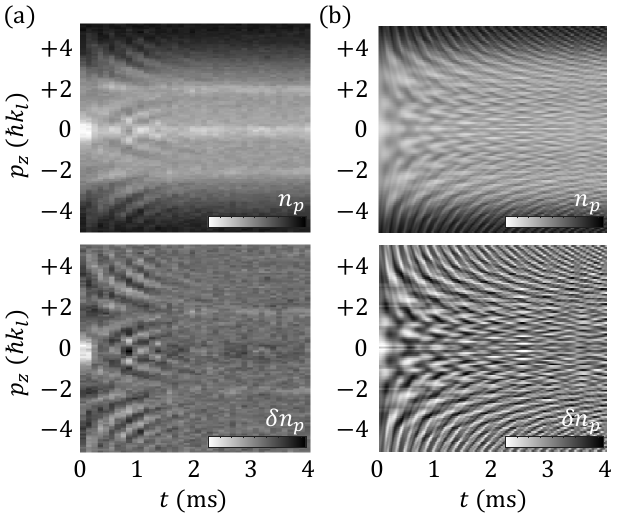}
\caption{
Calculations on interference patterns for thermal clouds in Michelson-type interferometers. (a) replicates data from Fig.~\ref{fig_temp}(d.i), for comparison with numerical simulations demonstrated in (b). The calculations capture the principal characteristics observed in the experimental data. 
}
\label{fig_simu}
\end{figure}

Interference phenomena are not exclusive to BECs, but are also present in thermal clouds~\cite{Miller2005, Sapiro2009}. To investigate the white-light characteristics of the implemented guided Michelson-type geometry, we make a modification to the experimental scheme used in Secs.~\ref{sec:MZ_experiment} and \ref{sec:MZ_interaction}. 
Interference is alternatively observed in the absence of a differential phase shift by varying the propagation time in the waveguide. We apply the recombining pulses at various times, (i) ranging from immediately after the splitting pulses to 4ms later, and (ii) within a time window of 8ms centered around the moment when the two packets are maximally overlapped upon their first return to the origin. 
The experimental sequences for these two groups of measurements are depicted at the top of Fig.~\ref{fig_temp}, with the labels (i) and (ii) used to indicate the respective experimental results. 
As shown in Fig.~\ref{fig_temp}(a), the measurements are performed on a nearly pure condensate with a scattering length of $487a_0$ and a particle number of $N=6000$. 
The interference is demonstrated by the oscillating occupation of particles in different momentum modes. The occupation in the modes $|0 \hbar k_l\rangle$ and $|\pm 2 \hbar k_l \rangle$ oscillates at a frequency of $4E_r/h$=1.0~kHz. 
Additionally, a minimal occupation is observed in the $| \pm 4\hbar k_l \rangle$ modes, which results from imperfections in the KD pulse sequence, oscillating at a fourfold-higher frequency of $16E_r/h$. 
The variations in the interference contrast over time are mainly due to the axial displacement of the two wave packets. The contrast reaches its (local) maximum when the two packets overlap at $t=0~$ms and again at half an oscillation ($t\approx 27~$ms), and diminishes as the packets gradually separate. 

Next, we perform measurements with clouds containing varying thermal populations by terminating evaporative cooling at various stages. We ensure that most atoms bind into molecules by evaporating them to temperatures significantly lower than their binding energy. The same lattice pulse sequences are applied for the interferometer operation. 
Figures~\ref{fig_temp}(b)-\ref{fig_temp}(d) illustrate the coherence properties of these thermal clouds, with each figure showing clouds at gradually increasing temperatures. The cloud shown in Fig.~\ref{fig_temp}(d) contains 95\% of molecules in the thermal component, with a temperature of 90nK. 
To better visualize the interference patterns, the deviation of the momentum distribution at each time from the time-averaged distribution, $\delta n_p(p_z,t)=n_p(p_z,t)-\langle n_p(p_z,t)\rangle _t$, is presented. The time carpet of $\delta n_p(p_z,t)$ is placed below each set of corresponding results. 

From these figures, we observe the presence of finer interference fringes in the cases of thermal clouds, superimposed on the interference signals from the condensates. With rising temperatures, the thermal clouds span a wider range of momentum modes, thereby extending the coherent fringe to a large momentum range of $\pm 10 \hbar k_l$. This observation illustrates the coherence imprinted by the beam splitter (optical lattice pulses). 
The emergence of these fringes results from the differential response of particles to lattice pulses across various momentum modes. During evolution, each momentum mode accumulates phase relative to its momentum. 
Notably, the presence of these coherent signals from thermal atoms remains distinctly visible even after half a period of oscillatory evolution in the waveguide, as demonstrated in Group (ii) in Fig.~\ref{fig_temp}. Our potential along the waveguide is very close to harmonic. From the design of the coils, we can calculate that the time difference in oscillation periods between the fastest and slowest particles in a thermal cloud is significantly smaller than the fringe width. This ensures the refocusing of all momentum modes at the precise time, which enables the observation of interference patterns after half an oscillation. 
The observation of interference fringes with thermal clouds signifies the white-light nature of the implemented Michelson-type interferometer and demonstrates the excellent robustness of our interferometric system. 

To support our interpretation of the fringes observed in the thermal clouds, we conduct a schematic numerical calculation to reproduce the interference pattern observed in Fig.~\ref{fig_temp}(d.i). We begin with a thermal cloud characterized by a Gaussian momentum distribution, with a $1/e^2$ width of $6\hbar k_l$, and segment it into discrete momentum modes based on the resolution of the experimental imaging. These modes are assumed to be mutually incoherent, with each treated as a plane wave denoted by $\lvert \hbar k_i\rangle$, where $i$ is the mode index. We calculate the state of the system following a sequence of lattice pulses [as given in Fig.~\ref{fig_osc}(a)], comprising four lattice pulses and three time intervals, totaling seven steps. This calculation is done using $\lvert\psi_{i,t}\rangle=\prod_{j=7}^1\hat{\mathcal{U}}_j\lvert \hbar k_i \rangle$. Here, $\hat{\mathcal{U}}_j=e^{-i[\hat{p}_z^2/(2m)+U_j\cos^2(k_l z)]t_j/\hbar}$ is the evolution operator in the $j^\mathrm{th}$ step. $t$ denotes the interference time, also referred to as $t_4$ in the calculations. $U_j$ equals zero during the intervals between pulses. The interaction term is neglected. 
By executing the evolution calculations individually across all momentum modes $\lvert \hbar k_i\rangle$ and then superimposing the outcomes, we obtain the interference fringes of thermal clouds. Varying the interference time $t$ in the calculations allows us to observe the interference pattern that emerges after the recombining pulses. The time step for the calculations is set to $20~\mathrm{\mu s}$, five times shorter than that used in the experiments ($100~\mathrm{\mu s}$). 
In Fig.~\ref{fig_simu}, we compare the experimental observations with the calculation results. This modeling approach replicates certain features of the experimental fringe patterns and captures the basic principles of the interferometer. Although these calculations do not encompass the full complexity of the experimentally measured patterns, they nonetheless provide valuable insights into the formation of the fringes, enhancing our understanding of the underlying physical processes.

\section{Conclusions}
\label{sec:conclusion}

Our studies investigated the performance of matter-wave interferometers using Feshbach molecules across a range of interaction strengths, extending into the very strongly interacting regime. Our experiments demonstrated controllability and readout of our molecular interferometers and provided an examination of the effects of interaction on signal properties. We observed and analyzed the influences on interference frequency, phase shift, and contrast, including its decay. Additionally, we explored particle loss due to scattering and the coherence of thermal particles. Where possible, we compared our findings with theoretical calculations, thereby establishing a quantitative and theoretical understanding of the results. Our work with extreme cases aims to be a further step towards understanding and potentially harnessing interactions for sensing and quantum mechanical devices.

\section*{Acknowledgements}
We thank Sebastian Erne and Igor Mazets for helpful discussions. 
This research was supported by the European Research Council: ERC-AdG ``Emergence in Quantum Physics'' (EmQ) under Grant Agreement No. 101097858 and the DFG/FWF CRC 1225 ``ISOQUANT'', with the Vienna participation financed by the Austrian Science Fund (FWF) [grant DOI: 10.55776/I4863].  This research was funded in whole or in part by the Austrian Science Fund (FWF) ``Non-equilibrium dynamics in strongly interacting 1D quantum systems'' [grant DOI: 10.55776/P35390]  and the ESPRIT grant ``Entangled Atom Pair Quantum Processor'' [grant DOI: 10.55776/ESP310]. For open access purposes, the author has applied a CC BY public copyright license to any author-accepted manuscript version arising from this submission. \\

\begin{appendix}
\section{Interferometer parameters}
\label{app:parameters}

In Table \ref{tab:quantities}, we provide an overview of the relevant quantities corresponding to the interferometers presented in Figs.~\ref{fig:ramsey_osc}(a)-\ref{fig:ramsey_osc}(g), \ref{fig_osc}(d), \ref{fig_aVar}(a), and \ref{fig_aVar}(b). These include the scattering length $a$ between molecules, total number of molecules $N$, and trap frequencies in the axial $\omega_z$ and radial $\omega_\perp$ directions. Additionally, we provide the chemical potential $\mu$, temperature $T$, binding energy $E_B$~\cite{Chin2010}, and mean-free path $\lambda$, which are calculated for the molecular condensate prior to the application of lattice pulses. 
The optical lattice is characterized by the lattice spacing $d$ and characteristic energy $E_r$. In the Ramsey-type interferometer experiments, the molecules are loaded into the optical lattice. The distribution of molecules across lattice sites results in a weighted-average molecule number per site $\bar N$. Furthermore, the collision-induced lifetime of molecules in $|d\rangle$ state $\tau_d$ is calculated for a cloud containing 50\% of molecules prepared in the $|d\rangle$ state~\cite{Zhai2013}. The tunneling time between the $|s\rangle$ states in neighboring lattice sites $\hbar/J$ is provided, and it is much longer than the time scale of our experiments. \\

\begin{table}[H]
    \centering
    \renewcommand{\arraystretch}{1.3} 
    \begin{tabular}{|c|c|c|c|c|c|c|c|c|c|c|}
    \hline
     \textbf{Fig.} & \textbf{2a} & \textbf{2b} & \textbf{2c} & \textbf{2d} & \textbf{2e} & \textbf{2f} & \textbf{2g} & \textbf{5d} & \textbf{6a} & \textbf{6b}\\
    \hline
    $a$ ($a_0$) & 333 & 487 & 693 & 982 & 1412 & 2109 & 3426 & 487 & 693 & 982\\
    \hline
    $N$ ($\times100$) & 35 & 75 & 85 & 95 & 95 & 90 & 75 & 60 & 60 & 60 \\ 
    \hline
    Trap freq.&  \multicolumn{10}{c|}{$\omega_z/2\pi\simeq$18.5~Hz, $\omega_\perp/2\pi\simeq$80~Hz} \\
    \hline
    $\mu/h$ (Hz)      & 214 & 338 & 409 & 491 & 568 & 653 & 737 & 309 & 356 & 409 \\
    \hline
    $k_BT/h$ (Hz)     & 350 & 453 & 471 & 488 & 488 & 480 & 453 & 419 & 419 & 419 \\
    \hline
    $E_B/h$ (kHz)    & 1946 & 910 & 449 & 224 & 108 & 49 & 18 & 910 & 449 & 224 \\
    \hline
    $\lambda$ ($\mathrm{\mu m}$)     & 223 & 97 & 56 & 33 & 20 & 12 & 6.3 & 106 & 64 & 40 \\
    \hline
    $d$ &  \multicolumn{10}{c|}{4.1~$\mathrm{\mu m}$} \\
    \hline
    $E_r/h$ &  \multicolumn{10}{c|}{249~Hz} \\
    \hline
    $\bar N$ & 320 & 516 & 556 & 556 & 509 & 451 & 358 & \multicolumn{3}{c|}{N/A} \\

    \hline
    $\tau_d$ (ms)    & 21 & 9.4 &  5.2 & 3.2 & 1.9 & 1.1  & 0.6 & \multicolumn{3}{c|}{N/A}\\
    \hline
    $\hbar /J$ & \multicolumn{7}{c|}{6600~s ($|s\rangle$ state)} & \multicolumn{3}{c|}{N/A}\\
    
    \hline

    \end{tabular}
    \caption{Relevant quantities to the interferometers as presented in Figs.~\ref{fig:ramsey_osc}(a)-\ref{fig:ramsey_osc}(g), \ref{fig_osc}(d), \ref{fig_aVar}(a), and \ref{fig_aVar}(b). }
    \label{tab:quantities}
\end{table}

\end{appendix}

\bibliography{Ref.bib}

\end{document}